\documentclass[12pt]{article}
\usepackage{fancyvrb,epsfig,amssymb,amsmath,bbm,psfrag,multirow}

\catcode`\@=11
\DeclareMathAlphabet{\mathpzc}{OT1}{pzc}{m}{it}

\font\cmss=cmss12 
\def\1{\hbox{{1}\kern-.25em\hbox{l}}}
\def\bfZ{\relax{\hbox{\cmss Z\kern-.4em Z}}}

\def \be  {\begin{equation}}
\def \ee  {\end{equation}}
\def \ba  {\begin{eqnarray}}
\def \ea  {\end{eqnarray}}
\def \baa {\begin{eqnarray*}}
\def \eaa {\end{eqnarray*}}
\def \bb  {\begin {thebibliography} }
\def \eb  {\end{thebibliography}}
\def \lab #1 {\label{#1}}

\def \matrix #1 {\left(\begin{array}{cc} #1 \end{array}\right)}

\newcommand{\as}{\ifmmode\alpha_{\rm s}\else{$\alpha_{\rm s}$}\fi}
\newcommand{\asbar}{\ifmmode\bar{\alpha}_{\rm s}\else{$\bar{\alpha}_{\rm s}$}\fi}

\newcommand{\ft}[2]{{\textstyle\frac{#1}{#2}}}

\newcommand{\Z}{{\mathbb Z}}

\font\cmss=cmss12 
\def\inbar{\,\vrule height1.5ex width.4pt depth0pt}
\def\IC{\relax\hbox{$\inbar\kern-.3em{\rm C}$}}
\def\IZ{\relax{\hbox{\cmss Z\kern-.4em Z}}}
\def\IR{{\hbox{{\rm I}\kern-.2em\hbox{\rm R}}}}

\def\IP{{\hbox{{\rm I}\kern-.2em\hbox{\rm P}}}}
\def\II{\hbox{{1}\kern-.25em\hbox{l}}}

\newbox\lett\newdimen\lheight\newdimen\lwidth
\def\ontop#1#2{\setbox\lett=\hbox{#2}\lheight\ht\lett
\multiply\lheight by 12 \divide\lheight by 10\relax%
\lwidth\wd\lett \multiply\lwidth by 8 \divide\lwidth by 10\relax #2\kern-
\lwidth%
\raise\lheight\hbox{{$\scriptstyle #1$}}\kern.1ex}

\def\XXint#1#2#3{{\setbox0=\hbox{$#1{#2#3}{\int}$}
     \vcenter{\hbox{$#2#3$}}\kern-.5\wd0}}

\begin{document}

\begin{titlepage}

\thispagestyle{empty}

\vspace*{1cm}

\centerline{\large \bf Realistic Four-Generation MSSM in Type II
String Theory}

\vspace{1cm}

\centerline{\sc A.V. Belitsky,  Richard F. Lebed, and Van E. Mayes}

\vspace{10mm}

\centerline{\it Department of Physics, Arizona State University}
\centerline{\it Tempe, AZ 85287-1504, USA}

\vspace{2cm}

\centerline{\bf Abstract}

\vspace{5mm}

We construct a four-generation MSSM with rank-4 Yukawa matrices from
intersecting D6 branes on a $T^6/(\Z_2 \times \Z_2)$ orientifold.  The
Yukawa matrices obtained provide an example of Flavor Democracy (FD),
where the Yukawa couplings are all nearly equal.  Mass hierarchies may
then be generated by slight perturbations away from FD.  We find that
it is possible to obtain hierarchical masses for the quarks and
leptons of each generation and mixings between them.  In addition, the
tree-level gauge couplings are unified at the string scale.  Finally,
we also construct similar models with one, two, and three generations
in which the rank of the Yukawa matrices is equal to the number of
generations in each model.

\end{titlepage}

\setcounter{footnote} 0

\newpage

\pagestyle{plain}
\setcounter{page} 1

{\bf 1. Introduction.}  The main challenge of string phenomenology is
to exhibit at least one string vacuum that describes the physics of
our universe in every detail.  Despite progress in this direction,
thus far this goal remains far from achieved.  In the past decade, a
promising approach to model building has emerged involving
compactifications with D branes on orientifolds (for reviews,
see~\cite{Ura03,BluCveLanShi05, BluKorLusSti06,Mar07}).  In such
models chiral fermions---an intrinsic feature of the Standard Model
(SM)---arise from configurations with D branes located at transversal
orbifold/conifold singularities~\cite{DouMoo96} and strings stretching
between D branes intersecting at
angles~\cite{BerDouLei96,AldFraIbaRabUra00} (or, in its T-dual
picture, with magnetized D
branes~\cite{Bac95,BluGorKorLus00,AngAntDudSag00}).  A number of
non-supersymmetric intersecting D-brane models have been constructed
that strongly resemble the SM.

However, non-supersymmetric low-energy limits of string theory suffer
from internal inconsistencies of noncanceled NS-NS tadpoles, yielding
models that destabilize the hierarchy of scales~\cite{CreIbaMar02A}.
A resolution of these issues necessarily requires $\mathcal{N} = 1$
supersymmetry.  The first semirealistic models that preserve the
latter were built in Type IIA theory on a $T^6 /(\Z_2 \times
\Z_2)$ orientifold~\cite{CveShiUra01a,CveShiUra01b}.  Subsequently,
intersecting D-brane models based on SM-like,
Pati-Salam~\cite{PatSal73}, and SU(5)~\cite{GeoGla74} gauge groups
were constructed within the same framework and systematically studied
in Refs.~\cite{CvePap06,CvePapShi02,CveLanLiLiu04,CheLiNan06}.  The
statistics of 3- and 4-generation models was studied
in~\cite{Blumenhagen:2004xx, Gmeiner:2005vz}.  Phenomenologically
interesting models have also been constructed on a $T^6/(\Z_6)$
orientifold~\cite{Honecker:2004kb}.  In addition, several different
models with flipped SU(5)~\cite{Bar81} have been suggested within
intersecting D-brane scenarios~\cite{EllKanNan02,AxeFloKok03}, as well
as models with interesting discrete-group flavor
structures~\cite{Abe:2009vi}.

Although F-theory model building~\cite{BeaHecVaf08} has received a
great deal of attention lately, intersecting D-brane models remain of
great interest.  While F theory is useful for overcoming the problem
of phenomenologically necessary but perturbatively forbidden Yukawa
couplings, one can also evade this problem in intersecting D-brane
models by utilizing the Pati-Salam gauge group ${\rm SU}(4) \times
{\rm SU}(2)_L \times {\rm SU}(2)_R$~\cite{CvePap06}, or by
incorporating nonperturbative effects such as D-brane
instantons~\cite{Abel:2006yk,IbaUra06,BluCveWei06}.  Furthermore, the
constraints that must be satisfied to construct globally consistent
intersecting D-brane models are well understood (see, {\it
e.g.},~\cite{BluKorLusSti06,Mar03}).  Thus, intersecting D-brane
constructions offer an exciting avenue for model-building,
particularly for building SM-like and left-right symmetric models
based on the Pati-Salam gauge group.

The SM exhibits an intricate pattern of mass hierarchies and mixings
between the different generations.  One challenge of any string
construction is to explain this structure.  Within the framework of
D-brane modeling it was demonstrated that the Yukawa matrices $Y_{abc}
\sim \exp(- A_{abc})$ arise from worldsheet areas $A_{abc}$ spanning D
branes (labeled by $a$, $b$, $c$) supporting fermions and Higgses at
their intersections~\cite{AldFraIbaRabUra00,Cremades:2003qj}.  This
pattern naturally encodes the hierarchy of Yukawa couplings.  However,
for most string constructions, Yukawa matrices are of rank one.  In
the case of D-brane models built on toroidal orientifolds, this result
can be traced to the fact that not all of the intersections at which
the SM fermions are localized occur on the same torus.  To date only
one three-generation model is known in which this problem has been
overcome~\cite{CheLiMayNan07}, and for which one can obtain mass
matrices for quarks and leptons that nearly reproduce experimental
values.  Additionally, this model exhibits automatic gauge coupling
unification at the string scale, and all extra matter can be
decoupled.  It should be commented that the rank-1 problem for
toroidal models can also potentially be solved by D-brane
instantons~\cite{Abel:2006yk, Cvetic:2009yh, Cvetic:2009ez}.  However,
the conditions for including these nonperturbative effects are very
constraining, and at present there are no concrete realizations in the
literature in which all constraints may be satisfied.

Although present high-energy experimental data supports just three
generations of chiral fermions, a fourth generation remains viable as
long as the mass of the extra neutrino $\nu'$ is larger than $\ft12
M_Z$, and the fourth-generation charged-fermion masses $m_{t^\prime}$,
$m_{b^\prime}$, and $m_{\tau^\prime}$ lie in the correct mass ranges
to avoid constraints from direct searches and precision electroweak
measurements~\cite{HolHouHurManSulUne09}.  Recent WMAP7 \cite {Wmap7}
analysis points towards a higher (than three) number of relativistic
neutrino species, $N_{\rm eff} = 4.34 ^{+0.86}_{-0.88}$, and
explanations have been proposed based on sterile neutrinos with masses
at sub-eV scales~\cite{KraLunSmi10} (see also Refs. \cite{Khl99} for earlier discussion
of astrophysical implications of the fourth generation).  Furthermore, the existence of a
fourth generation can introduce additional CP-violating
effects~\cite{Lunghi:2008aa,Soni:2008bc, Soni:2010xh,Lunghi:2010gv}
and can have a strong effect on Higgs boson searches at
colliders~\cite{Belotsky:2002ym}.  While not directly related to our
present focus, these considerations suggest possible manifestations of
a fourth generation of fermions.

Interestingly, the principle of Flavor Democracy
(FD)~\cite{HarHauWey78}, where the Yukawa couplings for quarks and
leptons are all nearly equal, appears to favor the existence of four
generations of chiral fermions. In this scenario, the observed mass
hierarchies are generated as a by-product of slight variations away
from FD, which may result from the internal geometry of the string
construction. For example, the down-type quark masses can be generated
naturally in the FD approach by a Yukawa matrix which is nearly rank
1, while up-type quark masses can be generated by a Yukawa matrix
which is nearly rank 2, corresponding to near degeneracies in the
Yukawa couplings.  Furthermore, in the FD approach a seesaw mechanism
is not necessary to obtain small neutrino masses, and the scenario of
three light neutrinos with a heavy fourth is naturally obtained by
considering small perturbations away from FD such that the Yukawa
matrix for neutrinos is nearly rank 1.  This construction is desirable
in intersecting D-brane models since a Majorana mass term for
right-handed neutrinos is perturbatively forbidden and can only be
obtained nonperturbatively with D-brane instantons.  In fact, all
Yukawa couplings for quarks and leptons can be allowed perturbatively in
intersecting D-brane models with a Pati-Salam gauge group, and in such
models a standard Majorana term of the form $W_M = M N N$ may not even
be generated through D-brane instantons since it is forbidden by
$U(1)_{B-L}$, which is gauged at the string scale.  If light, sterile
neutrinos turn out to exist, it may be an indication that nature does
not utilize the seesaw mechanism to achieve tiny neutrino
masses. Finally, FD may reduce the amount of fine-tuning necessary to
obtain hierarchical Yukawa couplings in these models since it is
natural for the Yukawa couplings to be nearly degenerate, with the
observed mass hierarchies being generated by slight departures from
this degeneracy.

In the following, we develop a four-generation model constructed from
intersecting D6 branes on a $T^6/(\Z_2 \times \Z_2)$ orientifold in
which one can obtain realistic mass matrices.  In particular, the
Yukawa matrices are rank 4, and one can obtain nontrivial masses for
each generation, with the mass spectrum over the generations being
naturally hierarchical.  Additionally, the MSSM gauge couplings are
unified at the string scale.  Also, the hidden-sector gauge groups
introduced to satisfy the Ramond-Ramond (RR) tadpole cancellation
conditions become confining at high energies.  Finally, we construct
similar models with one, two, and three generations of matter in which
the rank of the Yukawa matrices equals the number of generations.

{\bf 2. Rank-Four MSSM.} The configuration of D branes must obey a
number of conditions in order to be a consistent model of particle
physics.  First, the RR tadpoles vanish via the Gauss' law
cancellation condition for the sum of D-brane and cross-cap
RR-charges~\cite{BluKorLusSti06,GimPol96}:
\be
\label{RRtadpole}
\sum_{\alpha \in {\rm stacks}} N_\alpha (\pi_\alpha + \pi_{\alpha^*})
- 4 \pi_{\rm O6} = 0 \, ,
\ee
written in terms of the three-cycles $\pi_\alpha = (n^\alpha_1,
l^\alpha_1) \times (n^\alpha_2, l^\alpha_2) \times (n^\alpha_3,
2^{-\beta} l^\alpha_3)$ that wrap $(n^\alpha_j, m^\alpha_j)$ times the
fundamental cycles $([a_j], [b_j])$ of the factorizable six-torus $T^6
= \prod_{j=1}^3 T^2_{(j)}$.  Here, the first two two-tori are
rectangular: $l_j^\alpha = m_j^\alpha$ ($j = 1,2$), while the third
two-torus can be rectangular ($\beta \! = \! 0$), or tilted such that
$l_3^\alpha = 2 m_j^\alpha + n_j^\alpha$ and $\beta = 1$.  In the
T-dual picture the tilt of the third cycle $[a_3^\prime] = [a_3] +
\ft12 [b_3]$ corresponds to turning on a non-zero NS-NS two-form $B$
field. However, it becomes nondynamical under the requirement of its
invariance under the orientifold projection $\Omega
\mathcal{R}$~\cite{BluKorLusOtt01}. As a consequence, its flux can
admit only two discrete values, resulting in two discrete values for
$\beta$. Each two-torus possesses the complex structure modulus
$\chi_j = R^{(j)}_2/R^{(j)}_1$ built from its radii $R^{(j)}_1$ and
$R^{(j)}_2$.  $\mathcal{N}=1$ supersymmetry, which is favored for
reasons of underlying consistent low-energy theories of particle
physics as well as for stability of D-brane configurations, is
preserved by choosing the angles between the D-brane stacks and
orientifold planes to obey the
condition~\cite{CveShiUra01a,CveShiUra01b}
\be
\theta^\alpha_1 + \theta^\alpha_2 + \theta^\alpha_3 = 0 \ \mbox{mod} \,
2\pi \, ,
\ee
with $\theta^\alpha_j = \arctan (2^{- \beta_j} \chi_j
l^\alpha_j/n^\alpha_j)$ and $\beta_{1,2} = 0$ and $\beta_3 =
\beta$. This condition can be written in terms of wrapping numbers 
satisfying the two equations
\begin{eqnarray}
x_A \tilde{A}_a + x_B \tilde{B}_a + x_C \tilde{C}_a + x_D \tilde{D}_a = 0, \nonumber \\
A_a/x_A + B_a/x_B + C_a/x_C + D_a/x_D < 0 ,
\end{eqnarray}
where
\begin{eqnarray}
\tilde{A}_a = -l^1_a l^2_a l^3_a, \ \ \ \tilde{B}_a = l^1_a n^2_a n^3_a, \ \ \ \tilde{C}_a = n^1_a l^2_a n^3_a, \ \ \ \tilde{D}_a = n^1_a n^2_a l^3_a, \nonumber \\
 A_a = -n^1_a n^2_a n^3_a, \ \ \ B_a = n^1_a l^2_a l^3_a, \ \ \ C_a = l^1_a n^2_a l^3_a, \ \ \ D_a = l^1_a l^2_a n^3_a,
\end{eqnarray}
and $x_A$, $x_B$, $x_C$, and $x_D$ are the complex structure
parameters~\cite{CvePapShi02}, where $x_A = \lambda$, $x_B = \lambda
\cdot 2^{\beta_2 + \beta_3}/\chi_2 \chi_3$, $x_C = \lambda \cdot
2^{\beta_1 + \beta_3}/\chi_1 \chi_3$, $x_D = \lambda \cdot 2^{\beta_ +
\beta_2}/\chi_1 \chi_2$, and $\lambda$ is a positive parameter that
puts the parameters $A$, $B$, $C$, and $D$ on equal footing.
Furthermore, the consistency of the model is further ensured by the
K-theory conditions~\cite{Mar03,BluCveMarShi05}, which imply the
cancellation of the $Z_2$ charges carried by D branes in orientifold
compactifications in addition to the vanishing of the total
homological charge exhibited by Eq.\ (\ref{RRtadpole}). In the present
case, nonvanishing torsion charges are avoided by considering stacks
with an even number of D branes, {\it i.e.}, $N_\alpha \in 2
\mathbb{Z}$.

\begin{table}[t]
\caption{General spectrum for intersecting D6 branes at generic
angles, where $I_{aa'}=-2^{3-\beta}\prod_{i=1}^3(n_a^il_a^i)$, and
$I_{aO6}=2^{3-\beta}(-l_a^1l_a^2l_a^3
+l_a^1n_a^2n_a^3+n_a^1l_a^2n_a^3+n_a^1n_a^2l_a^3)$.  ${\cal M}$ is the
multiplicity, and $a_S$ and $a_A$ denote the symmetric and
antisymmetric representations of U$(N_a/2)$, respectively.}
\renewcommand{\arraystretch}{1.4}
\begin{center}
\begin{tabular}{|c|c|}
\hline {\bf Sector} & \phantom{more space inside this box}{\bf
Representation}
\phantom{more space inside this box} \\
\hline\hline
$aa$   & U$(N_a/2)$ vector multiplet  and 3 adjoint chiral multiplets  \\
\hline $ab+ba$   & $ {\cal M}(\frac{N_a}{2},
\frac{\overline{N_b}}{2})=
I_{ab}=2^{-\beta}\prod_{i=1}^3(n_a^il_b^i-n_b^il_a^i)$ \\
\hline $ab'+b'a$ & $ {\cal M}(\frac{N_a}{2},
\frac{N_b}{2})=I_{ab'}=-2^{-\beta}\prod_{i=1}^3(n_{a}^il_b^i+n_b^il_a^i)$ \\
\hline $aa'+a'a$ &  ${\cal M} (a_S)= \frac 12 (I_{aa'} - \frac 12
I_{aO6})$~;~~ ${\cal M} (a_A)=
\frac 12 (I_{aa'} + \frac 12 I_{aO6}) $ \\
\hline
\end{tabular}
\end{center}
\label{IBspectrum}
\end{table}

Imposing these constraints, we present the D6-brane configurations,
intersection numbers, and complex structure parameters of the model in
Table~\ref{MI-Numbers}, and the resulting spectrum in
Table~\ref{SpectrumB}, with formulas for calculating the multiplicity
of states in bifundamental, symmetric, and antisymmetric states shown
in Table~\ref{IBspectrum}.  Models with different numbers of
generations may be obtained for different values of the wrapping
number $n_g$ as well as the third-torus tilt parameter $\beta$.  The
observable sector of the models then has the gauge symmetry and matter
content of an $(N_g \! = \!  2^{1-\beta} n_g)$-generation SM with an
extended Higgs sector.  The extra matter in the models consists of
matter charged under the hidden-sector gauge groups, and vectorlike
matter between pairs of branes that do not intersect, as well as the
chiral adjoints associated with each stack of branes.  In addition,
one has matter in the symmetric triplet representation of SU(2)$_L$ as
well as additional singlets.  In order to have just the MSSM at low
energies, the gauge couplings must unify at some energy scale, and all
extra matter besides the MSSM states must become massive at
high-energy scales.  Furthermore, one requires just one pair of Higgs
doublets.

The resulting models have gauge symmetry $[{\rm U}(4)_C \times {\rm
U}(2)_L \times {\rm U} (2)_R]_{\rm observable} \times [ {\rm
USp}(2^{2-\beta}(4-n_g))^2 \times {\rm USp}(2^{2-\beta})^2]_{\rm
hidden}$. The hidden sector, as well as the set of complex structure parameters
required to preserve $\mathcal{N}=1$ supersymmetry, is different in each
of the models with
different numbers of generations.  In particular, in the tilted case
two of the hidden-sector gauge groups fall out in going from
three-generation to four-generation models. The non-Abelian chiral
anomalies vanish as a consequence of the RR tadpole condition
(\ref{RRtadpole}). The chiral anomalies from the three global U(1)s 
of U(4)$_C$, U(2)$_L$, and U(2)$_R$ inducing couplings of the form 
$A_\alpha \wedge F_\beta^2$, with $A$ and $F$ referring to Abelian and 
non-Abelian gauge fields, respectively, read~\cite{IbaMarRab01}
\be
\mathcal{A}^{\rm chiral}
= 
\ft12 \sum_{\alpha, \beta} N_\alpha (I_{\alpha\beta} -
I_{\alpha^\ast \beta})  A_\alpha \wedge F_\beta^2
\, ,
\ee
However, these
anomalies cancel against the couplings induced by RR fields via the
Green-Schwarz mechanism~\cite{IbaMarRab01}:
\be
\mathcal{A}^{\rm RR}
=
8 n_g A_a \wedge \left( F_c^2 - F_b^2 \right)
+
4 n_g A_b \wedge F_a^2 
-
4 n_g A_c \wedge F_a^2
\, ,
\ee
such that $\mathcal{A}^{\rm chiral} + \mathcal{A}^{\rm RR} = 0$. The
gauge fields $A_\alpha$ of these U(1)s receive masses via linear
$\sum_\ell c_\ell^\alpha B_2^\ell \wedge A_\alpha$ couplings in the
ten-dimensional action, with the massless modes given by ${\rm ker}
(c_\ell^\alpha)$.  The latter is trivial in the present model, which
means that the effective gauge symmetry of the observable sector is
${\rm SU}(4)_C \times {\rm SU} (2)_L \times {\rm SU}(2)_R$.

In order to break the gauge symmetry of the observable sector down to
the SM, we split the $a$ stack of D6 branes on the first two-torus
into stacks $a_1$ and $a_2$ with $N_{a_1}=6$ and $N_{a_2}=2$ D6
branes, and similarly split the $c$ stack of D6 branes into stacks
$c_1$ and $c_2$ such that $N_{c_1} = 2$ and $N_{c_2} = 2$.  The
process of brane-splitting corresponds to giving a vacuum expectation
value (VEV) to the chiral adjoint fields associated with each stack,
which are open-string moduli.  The gauge symmetry subsequently breaks
down to ${\rm SU}(3)_C \times {\rm SU}(2)_L \times {\rm U}(1)_{I_{3R}}
\times {\rm U}(1)_{B-L}$, where the U(1)$_{I_{3R}}$
and U(1)$_{B-L}$ gauge bosons remain massless.  The ${\rm
U}(1)_{I_{3R}} \times {\rm U}(1)_{B-L}$ gauge symmetry may then be
broken to ${\rm U}(1)_Y \! = \! \frac 1 2 {\rm U}(1)_{B-L} \!  + {\rm
U}(1)_{I_{3R}}$ by giving VEVs to the vectorlike particles with the
quantum numbers $({\bf 1}, {\bf 1}, 1/2, -1)$ and $({\bf 1}, {\bf 1},
-1/2, 1)$ under the ${\rm SU}(3)_C\times {\rm SU}(2)_L\times {\rm U}
(1)_{I_{3R}} \times {\rm U}(1)_{B-L}$ gauge symmetry arising from $a_2
c_1^\prime$ intersections. The full gauge symmetry of the models is
then ${\rm SU}(3)_C \times {\rm SU}(2)_L\times {\rm U}(1)_Y \times [
{\rm USp} [2^{2-\beta}(4-n_g)]^2 \times {\rm USp}(2^{2-\beta})^2]$,
with the hypercharge given by
\begin{equation}
Q_Y = \ft{1}{6} \left( Q_{a_1} - 3 Q_{a_2} - 3 Q_{c_1} + 3 Q_{c_2}
\right) \, ,
\end{equation}
where the $a$-stack charges provide $Q_{B-L}$ and the $c$-stack
charges provide $Q_{3R}$.  

\begin{table}[t]
\footnotesize
\renewcommand{\arraystretch}{1.3}
\caption{D6-brane configurations and intersection numbers for a series
of Pati-Salam models with $2^{1-\beta}n_g$ generations on a Type IIA
$T^6 / (\Z_2 \times \Z_2)$ orientifold, where the tadpole conditions
are satisfied without introducing fluxes.  The parameter $\beta$ can
be zero or one if the third torus is untilted or tilted respectively,
while the wrapping number $n_g$ may take the values $1$, $2$, $3$, or
$4$.  The complete gauge symmetry is $[{\rm U}(4)_C \times {\rm
U}(2)_L \times {\rm U}(2)_R]_{\rm observable} \times \left\{ {\rm
USp}[2^{2-\beta}(4-n_g)]^2 \times {\rm USp}
(2^{2-\beta})^2\right\}_{\rm hidden}$, and the complex structure
parameters that preserve $\mathcal{N}=1$ supersymmetry are $x_A =
x_B=n_g\cdot x_C = n_g \cdot x_D$.  The parameters $\beta^g_i$ give 
the $\beta$-functions for the hidden-sector gauge groups.}
\label{MI-Numbers}
\begin{center}
\begin{tabular}{|@{\,}c@{\,}||@{\,}c@{\,}|@{\,}c@{\,}||@{\,}c@{\,}|@{\,}c@{\,}|@{\,}c@{\,}|@{\,}c@{\,}|@{\,}c@{\,}|@{\,}c@{\,}|@{\,}c@{\,}|@{\,}c@{\,}|@{\,}c@{\,}|@{\,}c@{\,}|}
\hline
& \multicolumn{12}{c|}{${\rm U}(4)_C\times {\rm U}(2)_L\times
{\rm U}(2)_R \times {\rm USp}[2^{2-\beta}(4-n_g)]^2 \times {\rm USp}(2^{2-\beta})^2$}\\
\hline \hline  & $N$ & $(n^1,l^1)\times (n^2,l^2)\times

(n^3,l^3)$ 
& $n_{S}$& $n_{A}$ & $b$ & $b'$ & $c$ \ & $c'$& 1 \ & 2 \ & 3 \ & 4 \ \\
\hline
$a$&  8& $(0,-1)\times (1,1)\times (1,1)$ & 0 & 0  & $2^{1-\beta}n_g$ & 0 & $-2^{1-\beta}n_g$
& 0 & 1 & $-1$ & 0 & 0\\
$b$&  4& $(n_g,1)\times (1, 0)\times (1,-1)$ & $2^{1-\beta}(n_g-1)$ & $-2^{1-\beta}(n_g-1)$  & - & 0 & 0
& 0 & 0 & 1 & 0 & $-n_g$\\
$c$&  4& $(n_g,-1)\times (0,1)\times (1,-1)$ & $-2^{1-\beta}(n_g-1)$ & $2^{1-\beta}(n_g-1)$  & - & - &
- & 0 & $-1$ & 0 & $n_g$ & 0\\
\hline
1& $2^{2-\beta}(4-n_g)$& $(1,0)\times (1,0)\times (2^{\beta},0)$ & \multicolumn{10}{c|}
{$x_A = x_B = n_g \cdot x_C = n_g \cdot x_D \leftrightarrow \ \chi_1 = n_g, \ \ \chi_2 = 1, \ \ \chi_3=2^\beta$}\\
2& $2^{2-\beta}(4-n_g)$  & $(1,0)\times (0,-1)\times (0,2^{\beta})$ & \multicolumn{10}{c|}
{$\beta^g_1=-3,~\beta^g_2=-3$}\\
3& $2^{2-\beta}$ & $(0,-1)\times (1,0)\times (0,2^{\beta})$& \multicolumn{10}{c|}
{$\beta^g_3=-6+n_g$}\\    
4& $2^{2-\beta}$ & $(0,-1)\times (0,1)\times (2^{\beta},0)$& \multicolumn{10}{c|}
{$\beta^g_4=-6+n_g$}\\ 
\hline
\end{tabular}
\end{center}
\end{table}
\begin{table}
[htb] \footnotesize
\renewcommand{\arraystretch}{1.3}
\caption{The chiral and vectorlike superfields of the model, and their quantum
numbers under the gauge symmetry ${\rm U}(4)_C\times {\rm
U}(2)_L\times {\rm U}(2)_R \times {\rm USp}[2^{2-\beta}(4-n_g)]^2
\times {\rm USp} (2^{2-\beta})^2$.}
\label{SpectrumB}
\begin{center}
\begin{tabular}{|c||c||c||r|r|r||c|c|}\hline
& Multiplicity & Quantum Number & $Q_4$ & $Q_{2L}$ & $Q_{2R}$ & Field \\
\hline\hline
$ab$ & $2^{1-\beta}n_g$ & $(4,\overline{2},1,1,1,1,1)$ & 1 & $-1$ & 0  & $F_L(Q_L, L_L)$\\
$ac$ & $2^{1-\beta}n_g$ & $(\overline{4},1,2,1,1,1,1)$  & $-1$ & 0    & 1  &      $F_R(Q_R, L_R)$\\
$a1$ & 1              & $(4,1,1,\overline{N}_1,1,1,1)$ & 1 & 0 & 0  & $X_{a1}$\\
$a2$ & 1              & $(\overline{4},1,1,1,N_2,1,1)$ & $-1$ & 0 & 0   & $X_{a2}$ \\
$b2$ & 1              & $(1,2,1,1,\overline{N}_2,1,1)$ & 0 & 1 & 0    & $X_{b2}$ \\
$b4$ & $n_g$          & $(1,\overline{2},1,1,1,1,N_4)$ & 0 & $-1$ & 0    & $X_{b4}^i$ \\
$c1$ & 1              & $(1,1,\overline{2},N_1,1,1,1)$ & 0 & 0 & $-1$    & $X_{c1}$ \\
$c3$ & $n_g$          & $(1,1,2,1,1,\overline{N}_3,1)$ & 0 & 0 & 1   &  $X_{c3}^i$ \\
$b_{S}$ & $2^{1-\beta}(n_g-1)$ & $(1,3,1,1,1,1,1)$ & 0 & 2 & 0   &  $T_L^i$ \\
$b_{A}$ & $2^{1-\beta}(n_g-1)$ & $(1,\overline{1},1,1,1,1,1)$ & 0 & $-2$ & 0   & $S_L^i$ \\
$c_{S}$ & $2^{1-\beta}(n_g-1)$ & $(1,1,\overline{3},1,1,1,1)$ & 0 & 0 & $-2$   & $T_R^i$ \\
$c_{A}$ & $2^{1-\beta}(n_g-1)$ & $(1,1,1,1,1,1,1)$ & 0 & 0 & 2   & $S_R^i$ \\
\hline\hline
$ab'$   & $n_g$ & $(4,2,1,1,1,1,1)$ & 1 & 1 & 0  & \\
&        $n_g$ & $(\overline{4},\overline{2},1,1,1,1,1)$ & $-1$ & $-1$ & 0 & \\
\hline
$ac'$   & $n_g$ & $(4,1,2,1,1,1,1)$ & 1 & 0 & 1  & $\Phi_i$ \\
&        $n_g$ & $(\overline{4}, 1, \overline{2},1,1,1,1)$ & $-1$ & 0 & $-1$ & $\overline{\Phi}_i$\\
\hline
$bc$    & $2 n_g$ & $(1,2,\overline{2},1,1,1,1)$ & 0 & 1 & $-1$   & $H_u^i$, $H_d^i$\\
        & $2 n_g$ & $(1,\overline{2},2,1,1,1,1)$ & 0 & $-1$ & 1   &  \\
\hline
\end{tabular}
\end{center}
\end{table} 

The gauge coupling constant associated with a stack $\alpha$ is given
by
\begin{equation}
\label{idb:eq:gkf}
g_{{\rm D6}_\alpha}^{-2} = | \Re{\rm e} \,(f_\alpha) | 
\, , 
\end{equation}
where $f_\alpha$ is the holomorphic gauge kinetic function associated
with stack $\alpha$, given~\cite{BluKorLusSti06,CreIbaMar02A} in terms
of NS-NS fields by:
\begin{eqnarray}
f_\alpha &=& \frac{1}{4\kappa_\alpha}
\left[ n^\alpha_1\, n^\alpha_2\,n^\alpha_3 \,s 
- 
2^{-\beta} n^\alpha_1\,l^\alpha_2 \, l^\alpha_3 \, u^1 
- 
2^{-\beta} n^\alpha_2 \, l^\alpha_1 \, l^\alpha_3 \, u^2
- 
n^\alpha_3 \, l^\alpha_1 \, l^\alpha_2 \,u^3 \right] 
\, ,
\label{kingaugefun}
\end{eqnarray}
where $\kappa_\alpha = 1$ for SU($N_\alpha$) and $\kappa_\alpha = 2$
for USp($2N_\alpha$) or SO($2N_\alpha$) gauge groups.  The holomorphic
gauge kinetic function associated with SM hypercharge U(1)$_Y$ is then
given by taking a linear combination of the holomorphic kinetic gauge
functions from all of the stacks~\cite{BluLusSti03}:
\begin{equation}
f_Y = \ft{1}{6}f_{a_1} + \ft{1}{2} \left( f_{a_2} + f_{c_1} + f_{c_2}
\right).
\end{equation}
Note that in Eq.~(\ref{kingaugefun}), the four-dimensional dilaton $s$
and complex structure moduli $u^i$ refer to the supergravity basis.
These moduli must be stabilized, and gaugino condensation of the
effective Veneziano-Yankielowicz Lagrangian~\cite{VenYan82} provides
an example of such a mechanism~\cite{CveLanWan03}.  Gaugino
condensation in the hidden sectors can play an important role in
moduli stabilization, and it might provide a top-down reason why three
generations is preferred over four.

From the complex structure parameters, the complex structures $U^i$ are determined 
to be
\begin{equation}
U^1 = n_g \cdot i, \ \ \ \ \  U^2 = i, \ \ \ \ U^3 = -\beta + i.   
\end{equation}
The dilaton and complex structure moduli are then given in the supergravity basis by\footnote{See,
e.g., footnote 5 of Ref.\  \cite{BluLusSti03} for the relation between these and complex structures $U^i$.}
\begin{eqnarray}
\mbox{Re}(s) &=& \frac{1}{\left( 2^\beta n_g \right)^{1/2}} \frac{e^{-\phi_4}}{2\pi} \, , \ \ \ \ \
\mbox{Re}(u^1) \ = \left( \frac{2^\beta}{n_g} \right)^{1/2} \frac{e^{-\phi_4}}{2\pi} \, , \ \ \nonumber \\ 
\mbox{Re}(u^2) &=& \left( 2^\beta n_g \right)^{1/2} \frac{e^{-\phi_4}}{2\pi} \, , \ \ \
\mbox{Re}(u^3)  =  \frac{1}{\left( 2^\beta n_g \right)^{1/2}} \frac{e^{-\phi_4}}{2\pi} \, ,
\end{eqnarray}
where $\phi_4 = \ln g_s$ is the four-dimensional dilaton.  Inserting
these expressions into Eq.~(\ref{idb:eq:gkf}), one finds that the
gauge couplings are unified as $g^2_{s} = g^2_{w} = \frac{5}{3}g^2_Y =
g^2$ at the string scale $M_X$,
\begin{equation}
\frac{g^2 (M_X)}{4 \pi}
=
\left( \frac{2^\beta}{n_g} \right)^{1/2}e^{\phi_4}, 
\end{equation}
with the value of $\phi_4$ fixed by the value of the gauge couplings
where they unify, $g^2(M_X)$, which assumes different values for models
with different numbers of generations at $M_X =2.2\times 10^{16}$~GeV:
\be
g^2|_{N_{g} = 1}(M_X) = 0.275
\, , \
g^2|_{N_{g} = 2}(M_X) = 0.358
\, , \
g^2|_{N_{g} = 3}(M_X) = 0.511
\, , \
g^2|_{N_{g} = 4}(M_X) = 0.895
\, . 
\ee
The corresponding string scale is then given by
\begin{equation}
M_{\rm St} = \frac{g^2(M_X)}{4 \pi} \left( \frac{n_g \pi}{2^\beta}
\right)^{1/2} M_{\rm Planck},
\end{equation}
where $M_{\rm Planck}$ is the reduced Planck scale, $2.44 \times
10^{18}$~GeV.

After fixing the value of $\phi_4$, one can then determine the values
of the gauge couplings for the hidden-sector gauge groups at the
string scale:
\begin{equation}
g^2_{{\rm USp}_j} = 2^{(4-\beta/2)} \pi n_g^{(\rho_j/2)} e^{\phi_4},
\end{equation}
where $\rho_1 = \rho_2 = +1$ and $\rho_3 = \rho_4 = -1$.  Using the
beta-function parameters $\beta_j$ in Table~\ref{MI-Numbers}, the
scale at which each hidden-sector gauge group becomes confining can be
calculated:
\begin{equation}
\Lambda_j = M_X \cdot \mbox{exp} \left\{\frac{2\pi}{-\beta_j}
\left[1-\frac{2^{\beta}\pi}{g^2(M_X) n_g^{(\rho_j +1)/2}}\right]
\right\} \, .
\end{equation}

It can then be checked that the hidden-sector gauge groups have
sufficiently negative $\beta_j$ to become confining at high-energy
scales.  To have only one pair of light Higgs doublets, as is
necessary in the MSSM in order for the gauge couplings to unify, one must 
fine-tune the mixing parameters of the Higgs doublets,
specifically by fine-tuning the $\mu$ term in the superpotential,
which may be generated via the higher-dimensional
operators~\cite{CheLiMayNan07}:
\begin{eqnarray}
W & \supset & \frac{y^{ijkl}_{\mu}}{M_{\rm St}} S_L^i S_R^j
H_u^k H_d^l  ~,~\,
\label{eqn:HiggsSup}
\end{eqnarray}
where $y^{ijkl}_{\mu}$ are Yukawa couplings, $M_{\rm St}$ is the
string scale, and the singlets $S_R^j$ are assumed to receive
string-scale VEVs, while the VEVs of the singlets $S_L^i$ are
TeV-scale.  Note that this term may only be generated for the models
with $n_g > 1$.  The exact linear combinations that give the two light
Higgs eigenstates are correlated with the pattern of Higgs VEVs
necessary to obtain Yukawa matrices for the quarks and leptons,
\begin{equation}
H_{u,d} = \sum_i \frac{v^i_{u,d}}{\sqrt{\sum(v^i_{u,d})^2}},
\end{equation}
where $v^i_{u,d} = \left\langle H^i_{u,d} \right\rangle$.  Thus, at
low energies one obtains MSSM-like models with different numbers of
generations, with gauge-coupling unification $\sim 2.2 \times
10^{16}$~GeV, and matter charged under the hidden-sector gauge groups
becomes confined into massive bound states at high-energy scales.

As has been mentioned, quantities such as gauge and Yukawa couplings
depend on the VEVs of the closed-string moduli that parametrize the
size and shape of the compactified manifold, as well as the
open-string moduli that parametrize the positions of the D6-branes in
the internal space, which are associated with the presence of three
chiral adjoints in each stack.  These VEVs should be determined
dynamically.  While it is not our goal to solve this problem in the
present work, it should be mentioned that mechanisms do exist by which
this can be accomplished.  In particular, the closed-string moduli can
be stabilized in AdS by turning on fluxes in Type
IIA~\cite{Camara:2005dc}.  In fact, this mechanism has already been
demonstrated for the three-generation model~\cite{Chen:2006gd}.  Also,
gaugino condensation in the hidden sectors can provide another source
of closed-string moduli stabilization~\cite{CveLanWan03}.  The
open-string moduli may be frozen if the D-branes wrap rigid cycles, a
possibility that can exist on the $T^6/(\Z_2 \times \Z_2)$ orientifold
with discrete
torsion~\cite{Dudas:2005jx,Blumenhagen:2005tn,Forste:2010gw}.  An
example of a four-generation MSSM-like model constructed from
D6-branes wrapping rigid cycles is given in~\cite{CheLiNan06}.  We
emphasize the possibility of finding a dynamical reason to explain why
nature chooses a specific number of chiral generations by studying the
moduli stabilization problem for models with different numbers of
generations, such as our mini-landscape of models.

{\bf 3. Yukawa Couplings.} As one can see from the previous section
(note the filler brane stacks in Table~\ref{SpectrumB}), only the
models with $n_g \leq 4$ can satisfy the tadpole conditions without
introducing fluxes.  If we take this condition as a constraint, then
the only viable models from the top-down point of view have $N_g = 1,
2, 3, 4, 6$, and $8$.  Furthermore, masses may be generated via
trilinear couplings for all generations only for those models with a
tilted third torus ($\beta = 1$).  If we also take this condition as a
constraint, then the only viable models are those with $N_g = 1, 2,
3$, and $4$.  Additionally, the SU(3)$_C$ factor in the SM gauge group
is only asymptotically free for SUSY models with four generations or
less.  Thus, the maximum viable number of generations is four.

The three-generation model has previously been studied
in~\cite{CheLiMayNan07}.  As mentioned in the Introduction, this model
exhibits rank-3 Yukawa matrices and it is possible to nearly reproduce
the correct masses and mixings for the three known generations of
quarks and leptons.  However, since a fourth generation has not yet
been definitely ruled out, it is worth considering such models to see
if they can reproduce the observed masses and mixings for the known
quarks and leptons, while simultaneously satisfying experimental
constraints on the fourth generation.  In particular, the model with
$n_g = 4$ and a tilted third torus ($\beta = 1$) has rank-4 Yukawa
matrices, since all intersections occur on the first torus, similar to
the three-generation model.  In contrast to the three-generation
model, the Yukawa matrices in the four-generation model may
potentially be flavor democratic, and thus in some sense more natural.
Therefore, studying the Yukawa textures generated by this model is of
great interest.

A complete form for the Yukawa couplings $y^f_{ij}$ for D6-branes
wrapping on a full compact space $T^2 \times T^2 \times T^2$ can be
expressed as~\cite{CvePap06,Cremades:2003qj}:
\begin{equation} \label{Yukawas}
Y_{\{ijk\}}=h_{qu} \sigma_{abc} \prod_{r=1}^3 \vartheta
\left[\begin{array}{c} \delta^{(r)}\\ \phi^{(r)}
\end{array} \right] (\kappa^{(r)}),
\end{equation}
where
\begin{equation}
\vartheta \left[\begin{array}{c} \delta^{(r)}\\ \phi^{(r)}
\end{array} \right] (\kappa^{(r)})=\sum_{l \in\mathbf{Z}} e^{\pi
i(\delta^{(r)}+l )^2 \kappa^{(r)}} e^{2\pi i(\delta^{(r)}+l )
\phi^{(r)}},   \label{Dtheta}
\end{equation}
with $r=1,2,3$ denoting the three two-tori.  The input parameters are
given by
\begin{eqnarray} 
\nonumber
&&\delta^{(r)} = \frac{i^{(r)}}{I_{ab}^{(r)}} +
\frac{j^{(r)}}{I_{ca}^{(r)}} + \frac{k^{(r)}}{I_{bc}^{(r)}} +
\frac{d^{(r)} ( I_{ab}^{(r)} \epsilon_c^{(r)} + I_{ca}^{(r)}
\epsilon_b^{(r)} + I_{bc}^{(r)} \epsilon_a^{(r)}
)}{I_{ab}^{(r)} I_{bc}^{(r)} I_{ca}^{(r)}} +
\frac{s^{(r)}}{d^{(r)}}, \\ \nonumber &&\phi^{(r)} =
\frac{I_{bc}^{(r)} \theta_a^{(r)} + I_{ca}^{(r)}
\theta_b^{(r)} + I_{ab}^{(r)} \theta_c^{(r)}}{d^{(r)}}, \\ 
&&\kappa^{(r)} = \frac{J^{(r)}}{\alpha'} \frac{|I_{ab}^{(r)}
I_{bc}^{(r)} I_{ca}^{(r)}|}{(d^{(r)})^2}.
\label{eqn:Yinput}
\end{eqnarray}
where the indices $i^{(r)}$, $j^{(r)}$, and $k^{(r)}$ label the
intersections on the $r^{th}$ torus, $d^{(r)} \! = gcd(I^{(r)}_{ab},
I^{(r)}_{bc}, I^{(r)}_{ca})$, and the integer $s^{(r)}$ is a function
of $i^{(r)}$, $j^{(r)}$, and $k^{(r)}$ corresponding to different ways
of counting triplets of intersections.  The shift parameters
$\epsilon_a^{(r)}$, $\epsilon_b^{(r)}$, and $\epsilon_c^{(r)}$
correspond to the relative positions of stacks $a$, $b$, and $c$,
while the parameters $\theta_a^{(r)}$, $\theta_b^{(r)}$, and
$\theta_c^{(r)}$ are Wilson lines associated with these stacks.  For
simplicity, we set the Wilson lines to zero.  The brane shifts and
Wilson line together comprise the open-string moduli, which must be
stabilized in a complete model, as mentioned previously.  For the
present work, we treat them as free parameters, as our primary
interest is simply to see if one can obtain realistic mass matrices;
we discussed possible mechanisms for moduli stabilization in the
previous section. Note that although the above formulas for the Yukawa
couplings are for $T^6 = T^2\times T^2 \times T^2$, they may be extended to the
present case $T^6/(\Z_2 \times \Z_2)$ by including all of the orbifold images
in the analysis.  However, in the present case the cycles wrapped by the orbifold images of 
a stack of D-branes $a$ are
homologically identical to the original cycle wrapped by the stack $a$.  In addition, the intersection
numbers between the cycles defined on the orbifold turn out to be the same as the intersection
numbers between those on the ambient torus.  Thus, the above formulas for
$T^6 = T^2\times T^2 \times T^2$ may
be used without change on $T^6/(\Z_2 \times \Z_2)$.     

We focus only on the first torus, as the Yukawa couplings from the
second and third tori only produce an overall constant.  We label the
left-handed fields, right-handed fields, and Higgs fields with the
indices $i$, $j$, and $k$ respectively, which may assume the values
\begin{eqnarray}
i \in \left\{0, 1, 2, 3\right\}, \ \ \ \ \ j \in
\left\{0, 1, 2, 3\right\}, \ \ \ \ \ k \in
\left\{0, 1, 2, 3, 4, 5, 6, 7\right\}.
\end{eqnarray}
A trilinear Yukawa coupling occurs for a given set of indices that
satisfy the selection rule
\begin{equation}
i + j + k = 0 \ \mbox{mod} \ 4. 
\label{selrule} 
\end{equation}
The resulting rank-4 Yukawa matrices, for each of the $U$, $D$, and
$E$ sectors, have the form
\begin{eqnarray}
Y \sim 
\sum_{j=0,4}
\left(\begin{array}{llll}
Y_{00,j}\cdot v_j & Y_{01,3+j}\cdot v_{3+j} & Y_{02,2+j}\cdot v_{2+j} & Y_{03,1+j}\cdot v_{1+j}\\
Y_{10,3+j}\cdot v_{3+j} & Y_{11,2+j}\cdot v_{2+j} & Y_{12,1+j}\cdot v_{1+j} & Y_{13,j}\cdot v_j \\
Y_{20,2+j}\cdot v_{2+j} & Y_{21,1+j}\cdot v_{1+j} & Y_{22,j}\cdot v_j & Y_{23,3+j}\cdot v_{3+j} \\
Y_{30,1+j}\cdot v_{1+j} & Y_{31,j}\cdot v_j & Y_{32,3+j}\cdot v_{3+j} & Y_{33,
2+j}\cdot v_{2+j} \\
\end{array} \right) 
\, ,
\label{genYukmat}
\end{eqnarray}
where $v_k = \left\langle H_{k+1} \right\rangle$.  These Yukawa
matrices are rank 4, although they tend in practice to appear
numerically close to being matrices of lower rank, due to both the
periodic properties of
Eq.~(\ref{Dtheta}) and the selection rule Eq.~(\ref{selrule}).  FD
Yukawa matrices can therefore arise even when the Higgs VEVs are all
distinct and of the same order.  However, the degree to which FD is
broken depends upon differences between the Higgs VEVs as well as on
the open-string and K\a"ahler moduli, and so the detailed values of masses and
mixings vary sensitively with the particular values of these
parameters.  Nevertheless, the moduli are not fine-tuned {\it per se},
since they turn out to be quite different for the $U$ and $D/E$
sectors.  Note that the Yukawa matrices for the four-generation model
with $n_g=2, \beta=0$ are at most rank 2, for then the intersection
numbers on the first torus satisfy $gcd(I_{ab}, I_{ac}, I_{ca}) = 2$.
Thus, at most two generations can obtain distinct masses in that
particular variant of the model.  

{\bf 4. Numerical Analysis.} While the model described above supports
rank-4 Yukawa matrices, the question of whether one can in fact obtain
a phenomenologically suitable hierarchy of masses and mixings (which
are somewhat different in the $U$, $D$, and $E$ sectors) requires a
detailed numerical study. In particular, it is of interest to see if
the FD approach may be implemented. Strictly speaking, the observed
masses and CKM elements must undergo RGE evolution to the unification
scale $M_{\rm GUT}$ in order to be compared to model predictions.
This requirement introduces two significant difficulties: First, the
masses of the fourth-generation fermions, and their mixings with the
other three, are of course unknown; for purpose of discussion, we
simply take $m_{t^\prime} \! = \!  m_{b^\prime} \! = \!
m_{\tau^\prime} \! = \! 400$~GeV, a choice that satisfies all current
bounds~\cite{Nakamura:2010zzi}.  Indeed, the combination of direct
observation bounds, electroweak precision tests, and perturbative
unitarity constraints from heavy-fermion scattering amplitudes place
strong constraints on the possible masses, both lower and upper
limits~\cite{Dawson:2010jx}.  Second, a well-known problem of
four-generation models, both supersymmetric and not, is the presence
of several large Yukawa couplings (due to several fermions with
electroweak-scale masses) that generate runaway Yukawa couplings at
the TeV scale and above.  While attempts have been made to stabilize
the numerical evolution of the RGEs in 4-generation models up to
$M_{\rm GUT}$ by including new matter fields~({\it
e.g.},~\cite{Murdock:2008rx}), questions of the robustness of such
models remain.  For our purposes, we simply assume that {\em some\/}
supersymmetric model exists (see~\cite{Godbole:2009sy} for a related
discussion) in which the Yukawa couplings all remain suitably small
(from a perturbative viewpoint) to justify RGE evolution up to $M_{\rm
GUT}$, and take the values of the fourth-generation Yukawas at that
scale to be 400~GeV divided by the mass of the corresponding
third-generation fermion, while the Yukawas of the first three
generations of fermions are those from the 3-generation MSSM RGE
evolution to $M_{\rm GUT}$~\cite{Ross:2007az,Fusaoka:1998vc}.  While
such a fit is admittedly a hodgepodge from a phenomenological point of
view, its purpose is merely to provide a proof of principle for the
possibility of suitably hierarchical rank-4 Yukawas.

Specifically, since overall multiplicative factors in each Yukawa
matrix $Y_U$, $Y_D$, $Y_E$ remain undetermined [Eq.~(\ref{Yukawas})],
we fit to the fermion mass ratios $m^{\rm ratio}_{k\ell} \! \equiv \!
m_k/m_\ell$ in each sector and the independent hierarchical CKM
elements $V_{ud} \! \sim \! V_{cs} \! \gg V_{us}$.  To accommodate the
hierarchical nature of the masses, we choose $\ln (m^{\rm ratio,
fit}_{k\ell} / m^{\rm ratio, exp}_{k\ell})$ as the actual quantities
to be fit, which we allow to vary by a chosen multiplicative factor
(see below) in order to define a unit of $\chi^2$.  Since the CKM
elements are dimensionless, their analogous contributions to $\chi^2$
are $\ln (|V^{\rm fit}_{ij}/ V^{\rm exp}_{ij}|)$, so that the full
$\chi^2$ function reads:

\begin{equation} \label{chi2}
\chi^2 = \sum_{k\ell} \left[ \ln (m^{\rm ratio, fit}_{k\ell} /
m^{\rm ratio, exp}_{k\ell})/\ln (x_m) \right]^2
+ \sum_{ij} \left [ \ln (|V^{\rm fit}_{ij} / V^{\rm exp}_{ij}|) /
\ln(x_{v,ij}) \right]^2 \, ,
\end{equation}
where the $x_m$ and $x_{v, ij}$ indicate that a multiplicative
discrepancy by $x$ in any observable amounts to a unit of $\chi^2$. We
chose $x_m = 1.1$, {\it i.e.}, a 10$\%$ discrepancy, for mass ratios
with indices in Eq.\ (\ref{chi2}) ranging over $k\ell \! = \! tc, \,
cu, \, tt^\prime, \, bs, \, sd, \, bb^\prime, \, \tau\mu, \, \mu e, \,
\tau\tau^\prime$, and $x_{v,ij} = 1.05$ for the diagonal $ij \! = \!
ud, \, cs$ and $x_{v,ij} = 1.1$ for the off-diagonal $ij \! = \! us$
entries of the CKM matrix. Since all inputs are dimensionless, the fit
is sensitive only to the ratios of VEVs $v^{(F)}_j/v^{(F)}_0$, where
$j \! = \! 1, \ldots \, 7$, and $F$ refers to the appropriate Higgs
VEV (that of $H_U$ or $H_D$).  The relevant model parameters all
appear on the first torus; they are given by $\kappa^\prime \! \equiv
\!  i\kappa^{(1)}$ and the shift parameters $\epsilon_m$ [{\it i.e.},
the portions of $\delta^{(1)}$ in Eq.~(\ref{eqn:Yinput}) that are
independent of $i$, $j$, and $k$], with $m \! = \! 1, 2, 3$ denoting
the $U, D, E$ sectors, respectively:
\begin{equation}
\epsilon_m \equiv \frac 1 8 \left( \epsilon_{c,m}^{(1)} -
\epsilon_{b,m}^{(1)} -2 \epsilon_{a,m}^{(1)} + 2 s_m^{(1)} \right) \, .
\end{equation}

The numerical inputs and fit values of two sample fits are given in
Table~\ref{Fit}.  A few comments on the fits are in order: First, we
find that the minimization of $\chi^2$ drives the simulation towards
the vicinity of a stable minimum of the K\"ahler parameter $\kappa
\sim 3$ even when starting far away from it, {\it e.g.}, $\kappa_{\rm
init} \sim 6$.  Second, a similar stability of the minimum is observed
for the VEVs $v^{(U),(D)}_0$.  Third, we find that the VEVs obey the
naturalness requirement of being all of the same order, {\it i.e.},
their ratios satisfy $v^{(U,D)}_j/v^{(U,D)}_0 = O(1)$.  Numerical
simulations demonstrated the existence of a large number of solutions
with parameters densely populating regions around typical ones shown
in Table~\ref{Fit}, with very small deviations from the computed
values. These results fall roughly into two classes, with smaller and
larger mixing between the third and fourth generation of quarks: Note
the $V_{tb^\prime}$ entries in Table~\ref{Fit}. The first set of rows
gives typical VEV ratios generated by the minimization of $\chi^2$.
In the second set of rows, one sees that the size of $\chi^2$ is
driven by largely by discrepancies from the large fourth-generation
fermion masses ($\sim$ 400 GeV), while the mass ratios of the SM
fermions agree to a high level of accuracy.  The mixing between the
third and fourth generation is much smaller in the latter case. The
values for extra generations that we calculate from the model are
consistent with recent fits to flavor physics data of Ref.\
\cite{Alok:2010zj} for $m_{t^\prime} = 400$--$600 \ {\rm GeV}$,
$|V_{t^\prime b^\prime}| = 0.998 \pm 0.006$ and $|V_{t^\prime b}| =
0.07 \pm 0.08$.

The resulting down-quark Yukawa parameters using the input parameters
shown in the two columns on the left-hand side of Table~\ref{Fit} are

\begin{eqnarray}
Y_D \sim 
\left(\begin{array}{llll}
7.222 & 8.015 & 7.236 & 8.016\\
8.001 & 8.501 & 8.005 & 8.500 \\
7.236 & 8.016 & 7.222 & 8.015 \\
8.005 & 8.500 & 8.001 & 8.501 \\
\end{array} \right) 
\, ,
\end{eqnarray}
while the up-quark Yukawa matrix is given by 
\begin{eqnarray}
Y_U \sim
\left(\begin{array}{llll}
5.868 & 14.371 & 5.837 & 14.415\\
10.969 & 5.495 & 10.966 & 5.498 \\
5.837 & 14.415 & 5.868 & 14.371 \\
10.966 & 5.498 & 10.969 & 5.495 \\
\end{array} \right). 
\end{eqnarray}
As can be seen, these matrices exhibit near degeneracies such that
$Y_U$ is close to rank 2, resulting in two heavy quarks ($t^\prime$,
$t$) and two light quarks ($c$, $u$).  $Y_D$ is also nearly rank 2,
but (it turns out) is closer to rank 1 than $Y_U$, resulting in two
heavy quarks ($b^\prime$, $b$) and two lighter quarks ($s$, $d$), with
$b^\prime$ one being significantly heavier than the others.  As
discussed earlier, these near-degeneracies arise naturally due to the
selection rule Eq.~(\ref{selrule}) and the periodic properties of
Eq.~(\ref{Dtheta}).  However, it must be reiterated that particular
numerical values of parameters marking the departure from FD are
required to obtain these matrices, as seen from the choices of Higgs
VEVs, the open-string moduli parameters $\epsilon_{a,b,c}$, and the 
K\a"ahler modulus on the first torus.  A
complete model would of course provide a mechanism for stabilizing the
moduli to these values.  Alternately, some portion of the departure
from FD Yukawa matrices may instead arise from D-brane
instanton-induced couplings, rather than coming entirely from the
Higgs and moduli VEVs.

\begin{table}
\renewcommand{\arraystretch}{1.3}
\caption{Experimental input and sample fit values to the model.  Note
that we have fixed the values of the fourth-generation masses to be
$400$~GeV.}
\label{Fit}
\begin{center}
\begin{tabular}{|c|c|c||c|c|c|}
\hline
\multicolumn{3}{|l||}{
$\kappa^\prime =  3.2452096$, \quad \ \,
$\epsilon_1 = 0.203163427$
}
&
\multicolumn{3}{l||}{
$\kappa^\prime = 3.3338221$, \quad \ \,
$\epsilon_1 = 0.210962338$
}
\\
\multicolumn{3}{|l||}{
$\epsilon_2 = -0.090518048$, 
$\epsilon_3 = -0.090479810$
}
&
\multicolumn{3}{l||}{
$\epsilon_2 = -0.078384580$,
$\epsilon_3 = -0.079039780$
}
\\
\multicolumn{3}{|c||}{
$\chi^2 = 111$
}
&
\multicolumn{3}{c||}{
$\chi^2 = 142$
}
\\
\hline
$j$ & $v^{(D)}_j/v^{(D)}_0$ & $v^{(U)}_j/v^{(U)}_0$ &
$j$ & $v^{(D)}_j/v^{(D)}_0$ & $v^{(U)}_j/v^{(U)}_0$ \\
\hline
1 & 1.1484243 & 1.2146412 &
1 & 1.1638620 & 1.1950213 \\ 
2 & 0.9603816 & 0.7789428 &
2 & 0.9282357 & 0.7697829 \\ 
3 & 1.2306571 & 1.4632390 &
3 & 1.2033923 & 1.1610830 \\ 
4 & 1.2967893 & 0.7500008 & 
4 & 1.3064203 & 0.7459922 \\ 
5 & 1.1462177 & 2.6474058 &
5 & 1.1390428 & 1.2873315 \\
6 & 1.4202164 & 0.8490424 & 
6 & 1.4979862 & 0.8369313 \\ 
7 & 1.2432859 & 2.2181503 &
7 & 1.2831043 & 1.2145519 \\
\hline
$k\ell$ & $m^{\rm ratio, fit}_{k\ell}$ & $m^{\rm ratio, exp}_{k\ell}$ 
&
$k\ell$ & $m^{\rm ratio, fit}_{k\ell}$ & $m^{\rm ratio, exp}_{k\ell}$ \\
\hline
$tc$ & 246.1185 & 247.50 &
$tc$ & 247.1696 & 247.50 \\
$cu$ & 290.6168 & 290.60 &
$cu$ & 290.5549 & 290.60 \\
$t^\prime t$ & 2.7551  & 2.2857 &
$t^\prime t$ & 5.4376  & 2.2857 \\
$bs$ &  23.7721 & 35.700 &
$bs$ &  35.6756 & 35.700 \\
$sd$ & 20.0002 & 19.860 &
$sd$ & 19.8532 & 19.860 \\
$b^\prime b$ & 93.7180 & 80.000 &
$b^\prime b$ & 134.1494 & 80.000 \\
$\tau \mu$ & 23.7466 & 21.830 &
$\tau \mu$ & 21.8330 & 21.830 \\
$\mu e$ & 211.1299 & 211.10 &
$\mu e$ & 211.0802 & 211.10 \\
$\tau^\prime \tau$ & 93.7182 & 225.1 &
$\tau^\prime \tau$ & 134.1025 & 225.1 \\
\hline
$ij$ & $|V^{\rm fit}_{ij}|$ & $|V^{\rm exp}_{ij}|$ &
$ij$ & $|V^{\rm fit}_{ij}|$ & $|V^{\rm exp}_{ij}|$ \\
\hline
$ud$ & 0.9763 & 0.9754 &
$ud$ & 0.9759 & 0.9754 \\
$us$ & 0.2164 & 0.22050 &
$us$ & 0.2185 & 0.22050 \\
$cd$ & 0.2164 & 0.22030 &
$cd$ & 0.2185 & 0.22030 \\
$tb$ & 0.9815 & 0.9995 &
$tb$ & 0.9981 & 0.9995 \\
$t b^\prime$ & 0.1916 & - &
$t b^\prime$ & 0.0610 & - \\
$t^\prime b^\prime$ & 0.9815 & - &
$t^\prime b^\prime$ & 0.9981 & - \\
\hline
\end{tabular}
\end{center}
\end{table}

The numerical analysis shows that the fourth-generation masses are
naturally larger than those of the third generation.  Additionally, we
find a natural mass hierarchy between each generation.  However, the
model does not account for nonvanishing CKM elements between the
second and third generation of fermions without significant deviations
in mass ratios.  The nontrivial values for these elements can easily
be introduced by incorporating contributions to Yukawa matrices from
four-point correlation functions~\cite{Chen:2008rx}.  Finally, we note
that the above analysis provides a nice example of Flavor Democracy,
which makes this analysis more natural and less fine-tuned compared to
that of the three-generation model~\cite{CheLiMayNan07}.  

{\bf 5. Conclusion.} We have constructed a series of MSSM-like models
with different numbers of chiral fermion generations from intersecting
D6 branes on a $T^6/(\Z_2 \times \Z_2)$ orientifold.  Each of these
models satisfies all global consistency conditions, including tadpole
cancellation, K-theory constraints, and conditions for preserving
$\mathcal{N}=1$ supersymmetry. For each of the models, we also find
that the tree-level gauge couplings are unified at the string scale.
In addition, for the models constructed with one tilted two-torus, we
find that the rank of the Yukawa mass matrices for quarks and leptons
equals the number of generations.  Thus, distinct masses for each
generation and mixings between them may be generated.

The Yukawa mass matrices for the three-generation model have
previously been studied.  For this model it was found that it is
possible to generate mass hierarchies and mixings that nearly match
those that are observed~\cite{CheLiMayNan07}.  In the present work, we
studied the Yukawa matrices of the four-generation model and found
that accomplishing the same for the known three generations while
simultaneously satisfying constraints on fourth generation fermions
may be possible.  This conclusion, of course, comes with the caveat
that the problem of the evolution of large Yukawa couplings can be
ameliorated.  Finally, the obtained Yukawa matrices provide a nice
implementation of Flavor Democracy, and as such seem somewhat less
fine-tuned in comparison to the three-generation model.  In
particular, the up-type Yukawa matrix is almost rank 2, resulting in
two quarks with large masses and two with small masses.  The down-type
Yukawa matrix is nearly rank 1, resulting in one quark with a large
mass, and three quarks with smaller masses.

As commented earlier, the only model of the type under study for which
it is possible to generate masses and mixings for each generation are
those in which the third two-torus is tilted.  For this subset of
models, the maximum number of generations that can be accommodated
while simultaneously satisfying the tadpole constraints (without
including supergravity fluxes) is four.  Furthermore, it is known that
the maximum number of generations in a supersymmetric model for which
QCD is asymptotically free is also four.

Although disfavored, four-generation models are still presently viable
phenomenologically.  It is interesting that one can construct such a
``realistic'' model with four generations.  At present, the reason
that our universe seems to contain only three generations is unknown.
Experimentally, the Large Hadron Collider (LHC) should be able to
determine the question of the existence of the fourth generation
definitively. From the top-down point-of-view, string theory does not
yet appear to uniquely determine the number of generations.  However,
it may be possible to find dynamical reasons for singling out three
generations once the moduli stabilization question has been completely
addressed.  We leave this question for future work.

\paragraph{Acknowledgements.}
We thank Gary Shiu for reading the manuscript and for helpful advice,
and Tianjun Li for helpful discussions involving four-generation
models.  This work was supported by the National Science Foundation
under Grant No.\ PHY-0757394.
 

\end{document}